# PERSONAL SAFETY TRIGGERING SYSTEM ON ANDROID MOBILE PLATFORM

Ashokkumar Ramalingam[1], Prabhu Dorairaj[2] and Saranya Ramamoorthy[3]

grashok4u@gmail.com, prabhuhearts@gmail.com, saranpisces19@gmail.com

Department of Electrical Engineering, Blekinge Institute of Technology,
Karlskrona, Sweden

**ABSTRACT**

*Introduction of Smart phones redefined the usage of mobile phones in the communication world. Smart phones are equipped with various sophisticated features such as Wi-Fi, GPS navigation, high resolution camera, touch screen with broadband access which helps the mobile phone users to keep in touch with the modern world. Many of these features are primarily integrated with the mobile operating system which is out of reach to public, by which the users can't manipulate those features. Google came up with an innovative operation system termed as ANDROID, which is open system architecture with customizable third party development and debugging environment which helps the user's to manipulate the features and to create their own customizable applications.*

*In this paper, 'Emergency Based Remote Collateral Tracking System' application using Google's Android Mobile Platform is addressed. Emergency is divided into three categories: heart beat based emergency, security threats like personal safety and road accidents. This application is targeted to a person who is driving a vehicle. Heart rate monitoring device is integrated with our application to sense the heart beat of a person driving the vehicle and if there is any abnormalities in the heart beat, then our application performs a dual role. One in which, application uses a GPS to track the location information of the user and send those location information as a message via SMS, email and post it on Facebook wall Simultaneously, an emergency signal is sent to Arduino Microcontroller.*

*Road accidents are quite common, this application is also designed to detect the accident using the sensors in the Android Mobile. Security threat can occur anywhere, our application also answers for personal safety, when the user interacts with the application by pressing the button, then automatically the application generates the geographical information and sends that location information via SMS and email to a pre-stored emergency contact and the same information will be posted on user's Facebook wall. This application is written in JAVA programming language which runs on Eclipse Integrated Development Kit.*

## 1. INTRODUCTION

Instauration of mobile devices gave birth to lot of innovative technology, and exchanging information globally has become more prominent. Smart phones gave a new dimension to the usage of mobile phones for the users. Apart from basic functionality such as messaging, calling and cameras, smart phones laid a way to portray a personal computer. Not only the mobile phone looks newer, it's the operations system and the applications which are built to meet the various features of the hardware made difference.

The mobile phone has now become a major source of information device which can be seen almost in everyone's hand in the world. Mobile devices with computing process ability have been widely used to access network via mobile communication network. Different categories of application such as games, social networks, and health care are being developed to meet the user's requirements. Each mobile user is of unique kind, one wants to use the basic functionality of the smart phones, the other want to use the built in application, the most advanced user who wants to play with the hardware and to develop his own customizable application. To answer each kind of user, Google mustered up a groundbreaking product called as "ANDROID", which includes an open source operating system, middleware and a user-interface [1, 3].

## 2. HISTORY OF MOBILE OPERATING SYSTEM

Operating system is the heart of mobile devices, which controls and interacts with the mobile hardware. Similar precept to an operating system such as Windows, Mac OS and Linux, that controls the desktop or laptop. Device which runs on operating system are smart phones, PDA's and tablet computers.

Everyone wants to do everything fast and on the go. When people where sitting back and diddling with the heavy computers for accessing the internet. An operating system called Palm OS was launched in year 1996 which brought a drastic change in the communication world [2]. With the introduction of Palm OS 2.0 in the year 1997, accessing and sending mail via mobile evolved. The time when Palm OS was standing alone in the Smart phone market in the year 2000, another giant bounced into the market, introducing Windows "Pocket PC 2000" which almost had most of the features of a computer.

Entertainment on the go was achievable with windows by launching "Pocket PC 2002" which incorporated MSN messenger and media player with enhanced user interface. Bluetooth an extraordinary invention for file transfer wirelessly. Bluetooth integration was successfully implemented in Windows Mobile 2003 and browsing was made more comfortable with the pocket internet explorer. When windows were acquiring the smart phones market, Palm OS Cobalt bounced back with wifi and Bluetooth connectivity in 2004.

In 2005, Google acquired the Android Inc and Blackberry's OS 4.1 was made available in the market. Windows interfaced the GPS management and office mobile in their "windows mobile 5". When everyone was going upwards in updating the version and integrating application in the smart phones. The release of "iPhone" in 2007 disrupted the mobile industry and gave a new era of smart phone operating system with user experience which relies on touch based user interaction.

In 2007, a trendsetting year when Google formed the OHA [4] with 79 other hardware, software and telecommunication companies to make entry in to the smart phone market by introducing a legendary open source operating system "ANDROID" resulted in 2008 with Android 1.0 which was available in the market. Android came up with a middleware which is responsible for hardware and communication between applications, and provides open source Android SDK application that allows embedded systems developers to use it to develop their own customizable Android platform applications. Some notable top applications such as Google map, E-mail, Instant messaging, Browser, GPS, Multimedia messaging are widely made available to the people in large only because of Android.

The enhancing grandness of smart phones has sparked off intense contenders amongst software giants such as Google, Microsoft, and Apple, as well as mobile industry leaders Nokia, RIM, and Palm to keep on updating their technology. In 2009, Samsung too joined the roads of smart phones when they released a new operating system called as BADA platform. Nevertheless hewlett packard Web OS was also introduced in the same year. But Google's Android was climbing so high in a year, they acquired the major share in the smart phone operating system by upgrading from Android 1.0, 1.1-1.6 till 2.1 (Éclair) and version 3.1 (Honeycomb) was released in 2011 [2].

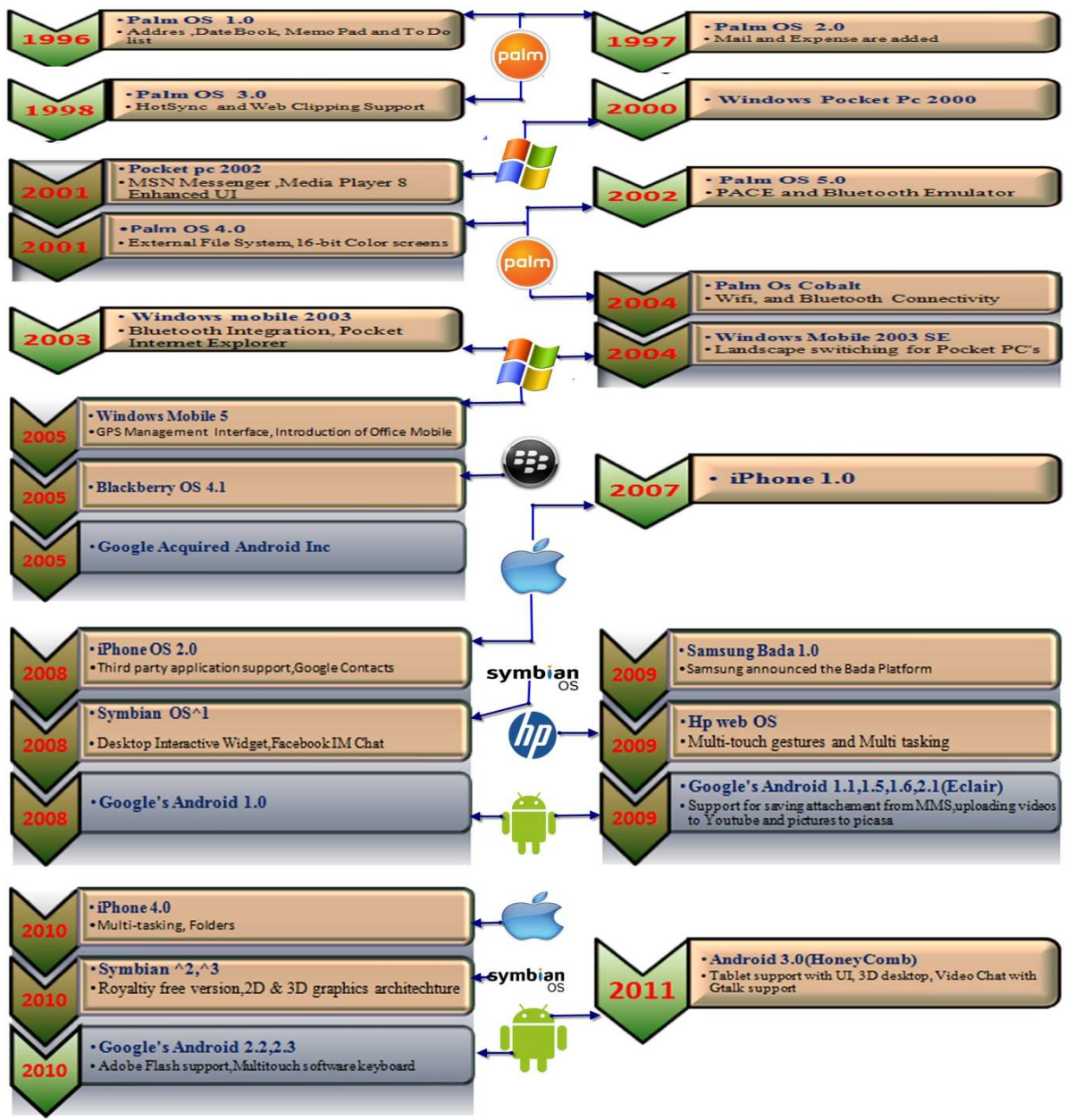

Figure 1. History of Mobile Operating System

## 3. WHAT IS ANDROID?

In 2005, Google acquired Android from Android Inc. which was found in year 2003 by Andy Rubin and they dealt with developing software for mobile devices. Later, OHA which comprises of 79 companies along with Google developed their new mobile platform for mobile devices. This alliance was formed so as to develop open technologies for mobile devices and make those applications easily available in the market. This new open source technology was named as **Android** [3, 4].

Android is an open source architecture which is used for developing applications for mobile devices. Android works on Linux Kernel. It has an operating system, middleware and key applications. Android announced its code under the license of free software/open source in the year 2008. Android comes up with an API for mobile devices. This Linux Kernel supports Java Virtual Machine which favours Java to be most suitable programming Language for development of the applications. Google provides a SDK to all developers which include libraries, debugger and a handset emulator in Eclipse IDE [5, 6]. The application which is developed in Android can be tested using this emulator which works similar to a mobile phone.

### 3.1 Arduino Microcontroller

Arduino Microcontroller is an open source prototyping platform which can sense the environment by the sensors which is given as the input to it. The programming of the controller is done using Arduino Programming language. The language used for programming is C/C++. Various Arduino microcontrollers are available in market such as Arduino Extreme, Arduino Mini, Arduino Nano, Arduino Bluetooth, Arduino Diecimila, Arduino Duemilanove, Arduino Mega and so on. Each of these microcontrollers have their own significance. Arduino Bluetooth is found as best choice for our project. As the name suggests, this microcontroller has in built Bluetooth module which lacks in other controllers [9].

Arduino Bluetooth (Arduino BT) microcontroller as in Fig.2 works on principle of Atmega168 and the Bluegiga WT11 Bluetooth module. It has 14 input/output pins and it supports serial communication over Bluetooth. The operating voltage is 5V which makes the controller very fragile and hence the voltage should not be exceeded else it would result in the damage of the microcontroller. It has 16kb flash memory for the storage of the code. The reset option is at pin number 7 which is connected to the reset of bluegiga WT11 module. The Bluetooth communication is provided by Bluegiga WT11 module on Arduino BT which can connect to any devices which has Bluetooth connectivity. It should be configured and should be detected by the device to which it is connected. It works on the baud rate of 112500. The controller is connected to another device by pairing and the name of the device suggested by Arduino is ARDUINOBT and the passcode is 12345. This is the default setting of the device [9]. Arduino Bluetooth microcontroller is connected to the Android mobile device via Bluetooth.

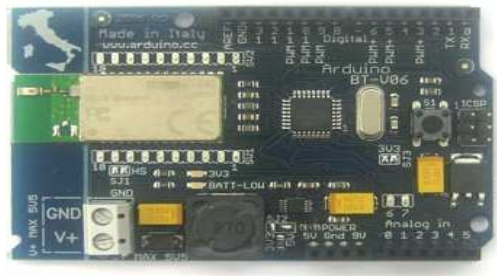

Figure 2. Arduino Microcontroller

The challenge is the compatibility of the microcontroller to the mobile phone enabled with android. The programming of the android mobile device should work well with arduino BT microcontroller which is proper integration of them. This amalgamation of the components is done using Amarino.

### 3.2 Heart Rate Monitoring Device

Heart is the main organ in a human's body. One can't live without it. It's because of this anything in the world is compared to heart, that's the importance of such a vital organ. Heart rate is an important factor to be considered in a human body. Heart rate tells us how many times heart beats in a minute. It is usually measured by feeling the pulse on any area near the artery. This measure signifies the blood pressure of a person. The blood pressure either low or high is dangerous to health. Hence it has to be kept under control and also by constant monitoring [10].

Heart rate monitoring is an important aspect of a human being. This monitoring is usually done by a regular health checkup at any hospital. This is a normal scenario, but there are situations where the heart rate is not monitored while driving any vehicle or while exercising and so on. Hence a heart rate monitoring device is very essential. Heart rate monitoring is done at any hospitals using devices like ECG. Even though it is accurate, this device is costly and also regular visit to hospital should also be carried out. Also a person with heart disease complaint should be able to monitor his condition continuously. To solve all these criticalities, a heart rate monitoring device has to be purchased and maintained for personal care. These days heart rate monitor is been used commonly by normal person rather than in a hospital [11, 12].

Heart rate monitor helps to detect the abnormalities in the heart and would display it to the person who is using it. This feature has inspired us to use such a device for our emergency conditions especially when a person is driving a car and is suddenly met with heart attack. To get situation under control, this monitor device would send an alert to the android enabled mobile phone which will in turn halt the car to avoid further causalities. There are various heart rate monitoring device are available in the market such as Zephyr HR Bluetooth heart rate monitor, polar Bluetooth heart rate monitor, Wahoo Fitness ANT plus Dongle and so on [13].

Zephyr HR Bluetooth enabled heart rate monitor as shown in Fig.3. is best suited for our project for various reasons such as the Zephyr programming is easier and it is open source. It is a device with Bluetooth connectivity which avoids wired connection and reduces the hardware cost for it. It also has a fabric sensor which detects the data irrespective of any fabric. Speed, distance is also displayed using this device which helps to see a pictorial representation of a person's heart rate. The best thing about this device is that it can tolerate any extreme motion of the body like running, jumping, jogging and so on. Also the transfer of data is via Bluetooth [13].

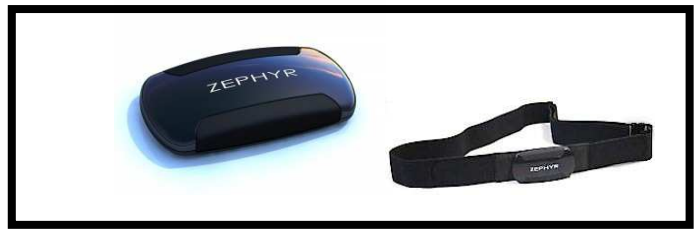

Figure 3.Zephyr Bluetooth Heart Rate Device

### 3.3 Integration of Arduino with Android-Amarino

Every request sent has its own response. The same is the case with a mobile phone. For instance, a phone call is alerted to the user by a ringtone, a text message received is displayed on the screen, a photo clicked with the help of the camera is saved in memory of the phone and so on. These events are generated on the phone itself. The same event can also be viewed somewhere else like in our room, through a sensor like accelerometer or on a microcontroller. To such a situation to occur, Amarino is used.

Amarino is a tool kit which helps in integration of android with arduino. It consists of the Android application and libraries required for arduino. Amarino helps to connect a mobile device enabled with android and an arduino microcontroller via Bluetooth [14].

## 4. EMERGENCY SCENARIO

### 4.1 Heart Rate Based

#### 4.1.1 Problem Statement

Nowadays, mobile devices started to integrate with various third party hardware's to provide more functionality to the users, which also leads to the integration of a heart rate device which will monitor the heart beat of an user. But how the heart rate device can be integrated with Android mobile, so that the android enable mobile can monitor the heart beat of a person, and also how to use that heart rate to manipulate a person under emergency? The Zephyr heart rate monitoring device is used to fetch the heart rate of a person and that device is integrated with the android mobile with help of programmable application, developed using android SDK and, this application will decide about the critical situation with respect to heart rate and sends a message to a pre-stored emergency contact number which also contains the geographical location of the user.

### 4.1.2 Scenario

The design of this project deals with a person/user driving in a vehicle. A typical scenario is, when a person driving in isolated roads, wearing the Zephyr heart rate device around the chest. This heart rate device will send the heart rate every second to the android mobile via Bluetooth by which it is monitoring the heart beat of the driving person. The heart rate is normal between 60-100, if it is less than 60 it's called has *Bradycardia* and if it's more than 100 it is called as *tachycardia*. In most condition the heart beat becomes less when there is dehydration, decreased protein intake and it becomes more in uncontrolled hypertension.

### 4.1.3 Modeling

Our application is designed to sense this heart rate, and if there are any abnormalities in the heart rate like, if the heart rate goes below 60 or above 100, automatically the android mobile will send a signal to an Arudino microcontroller which is connected to android mobile via Bluetooth. This Arduino microcontroller will make an alert signal, in our case the alert signal is indicated by blink of a led. Simultaneously our application will track the location information of the user who is under emergency and send that location information to a remote pre-stored emergency contact number. This scenario is shown in Fig.4.

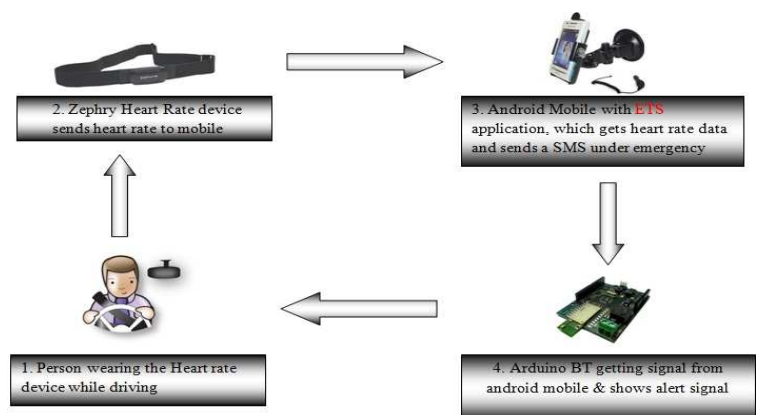

Figure 4.Flow diagram of Heart Rate Based Scenario

### 4.1.4 Implementation and Validation

#### 4.1.4.1. Implementation of API with Heart Rate Device

Implementation of communication between a zephyr heart rate (HR) device and android mobile starts with designing an API. Fig.3below shows the Zephyr Bluetooth communicates with a mobile device over the Bluetooth link. The Zephyr Bluetooth HR device uses a Bluetooth SPP (Serial Port Profile) to communicate with the low level protocol such as

a) 115,200 baud rate b) 8 data bits c) 1 stop parity bits d) No parity

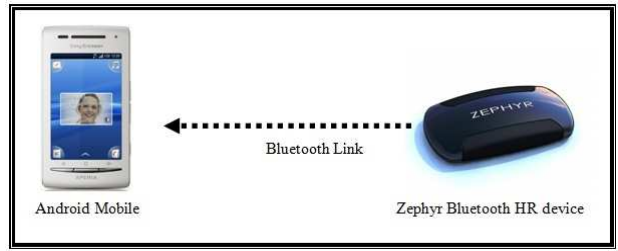

Figure 5. Zephyr Bluetooth device communication with Android phone

Our application employs the API and enables the HR device to transmit the different packet types such as heart rate, speed and distance packets. The following steps below are a description of the most important aspects of the source code in our application used to enable the General Packet and display the data on the Android phone.

1. On clicking the Connect button, a Bluetooth adaptor type object is created and passed to an Object of the *BT Client* class type. The *BT Client* object is essentially a thread that manages the overall Bluetooth connectivity of the phone with the HR device.

2. Next, an object of the *NewConnectedListener* class will need to be created which essentially implements the *ConnectedListener* interface, and one that extends the *ConnectedListener* class. This object is responsible for reacting differently to different kinds of messages. In this object we override the parent class's *connected* method and define our own method. In this method we create a *ZephyrProtocol* object and call its *addZephyrPacketEventListener* method. This method takes a *ZephyrPacketListener* argument, in whose *ReceivedPacket* method we define what message we are interested in, and how we want the data to be displayed on the phone screen.

3. This *ConnectedListenerImpl* object needs to then be connected to the *BTClient* object type via *addConnectedEventListener* function call to tie this object to respond to a received packet from the HR device.

4. Calling the *start* function of the *BTClient* thread kicks off the communication of the Application with the HR device.

**Validation:**

The HR device powers on automatically when worn. If there is insufficient skin conductivity (excessively dry skin and/or strap sensor pads), the wear-detect circuitry may not trip. Moisten skin and sensor pads with water. The heart rate device has to be worn around the chest region just near the sternum as shown in the Fig.6.

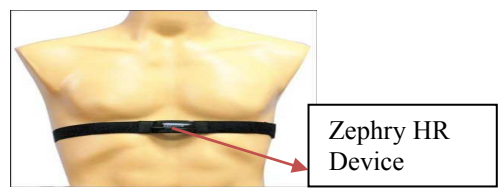

Figure 6. Zephyr heart rate device worn

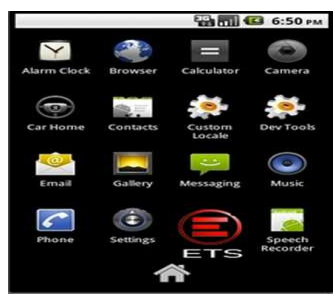

Figure 7. ETS Application

Open ETS-Emergency Tracking System application which is installed in the android mobile as shown in Fig.7. As soon as the ETS is clicked, home page with menus such as Pair HR Device, Pair Controller, Enter Details, Start and Disconnect is opened. In order to get the heart rate, user need to pair the HR device with the android mobile. As soon as the HR device is paired with the mobile, the heart rate starts to get displayed on the mobile screen. The entire workflow can be seen in the Fig.8.

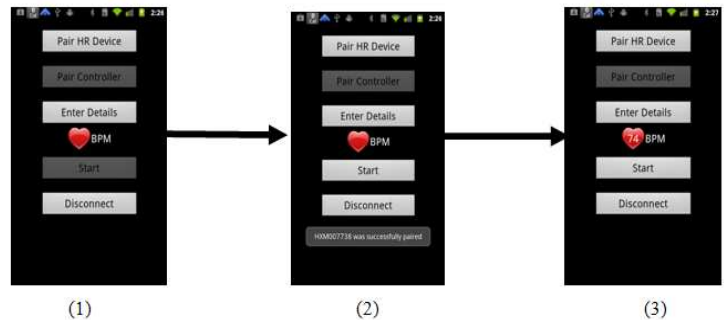

Figure 8. Entire Workflow of Pairing HR device to Android phone

1. Connects to HR device by tapping on "Pair HR Device"

2. A message shows HR device is successfully paired

3. Heart rate get displayed on the screen "74 BPM (beats per minute)"

The emergency contact number, email address and Facebook account can be integrated in the "Enter Detail page" in the main menu of ETS as shown in Fig.9.

Figure 9.Enter Details page

When HR device starts to transmit the heart rate to mobile, ETS application performs condition check with respect to 60<HR<120, if heart rate is less than 60 or more than 120 then the application decides that this condition is critical and starts to track the location of the user using GPS API and simultaneously sends a message to pre-stored emergency contact number and also to a pre-stored Email address. Finally a message containing "This person is under emergency take necessary action" followed by the geographical location of the person is posted on the enabled Facebook wall. Simultaneously an alert signal is send to arduino microcontroller, to acknowledge the risk signal a LED is connected to pin 13 of the microcontroller and that LED blinks under critical situation. The entire work flow is represented in Fig.10.

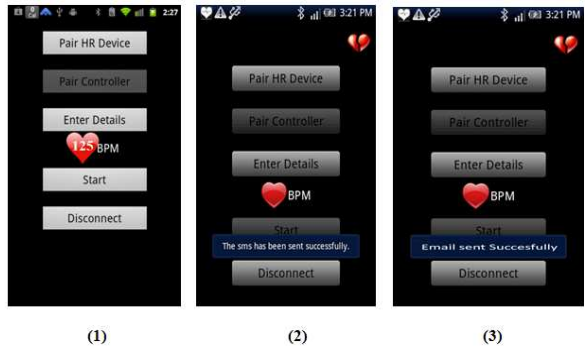

(1)                      (2)                    (3)

Figure 10. Abnormal heart rate scenario – workflow

**(1)** ETS reads an abnormal heart rate of 125BPM

**(2)** SMS was sent to a pre-stored number

**(3)** Email was sent to a pre-stored Email address

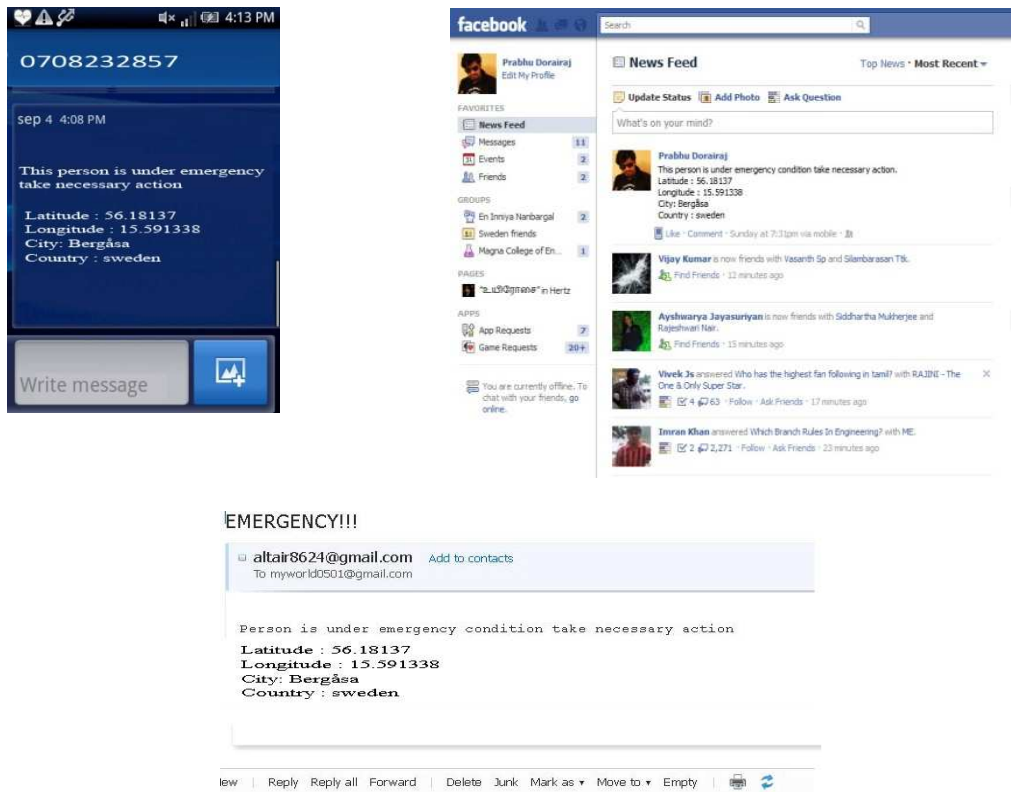

Figure 11. Message sent via SMS, E-Mail and Facebook Wall

## 4.2 Security Threats

### 4.2.1 Problem Statement

The use of GPS on mobile devices in the recent generations of mobile communication is one of the ubiquitous applications that are effectively developed. Symbian is the most utilized mobile operating system in this regard that most mobile devices make use of. Window's mobile, Apple's iPhone, Blackberry and others also have their own versions. But how can the location of a person be tracked and notified using Google's Android platform? The GPS Application Program Interface (API) is used to fetch the latitude and longitude coordinates and these are used by SMS manager to send the information as a SMS to a friend or emergency contact which is a pre-stored number. Thus, the rich application features of the Android like the GPS API and SMS manager can be used to track and notify.

This project is developed with Java programming language which is executed on eclipse IDE and implemented on Android Mobile Platform in turn the validation will be performed using an emulator. A simple graphical user interface is used to manage the scenario and make it eventually available to the public at large.

### 4.2.2 Scenario

Security threat can occur anywhere at any time. This project is designed to overcome personal security threat under user request. If the user is in emergency and if the user feels that his/her location has to be sent to a remote user then this project comes in to action by providing a exact geographical location information, which can be sent to pre-stored number and to E-mail address under user request.

### 4.2.3 Modeling

The model of this project is presented in Fig.12. It depicts the working model of this project to show the location-based mobile service design on Android and the SMS functionality as well. In order to fetch the current position of the user it is needed to obtain the longitude and latitude coordinate values. All the applications should carry an AndroidManifest.xml file (with precisely that name) in its root directory, since the essential information about the application will be held by the manifest. Hence we can access the protected part of APIs. To expend the GPS functionality, we add ACCESS_FINE_LOCATION to get permission to Androidmanifest.xml. To obtain the coordinate values, a location manager has to be created. Location manager is the part that is responsible for creating a location based service on Android. The snippet code below shows how the location manager can be created

*LocationManager lm = (LocationManager)*

*getSystemService(Context.LOCATION_SERVICE).*

After fetching the longitude and latitude coordinate values from GPS, a SMS message which contains the location information is generated. In order to send a SMS message, *SEND_SMS* permission is added to the AndroidManifest.xml file which in turn makes the application to send SMS message using *sendSMS()* function thereby initiating the *smsManager* to allow the application to send a SMS message.

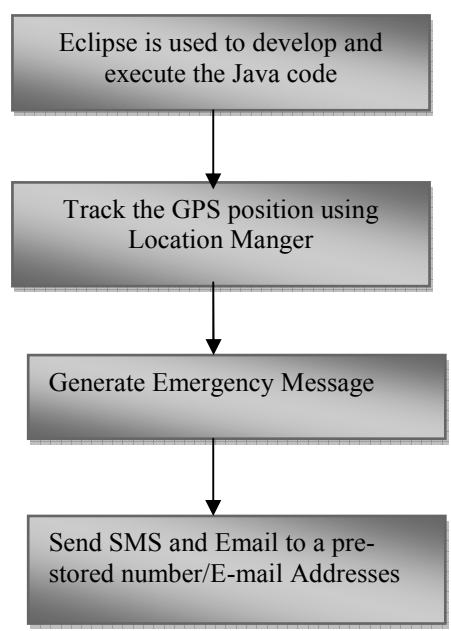

Figure 12.Modelling of Personal safety Scenario

### 4.2.4 Implementation and Validation

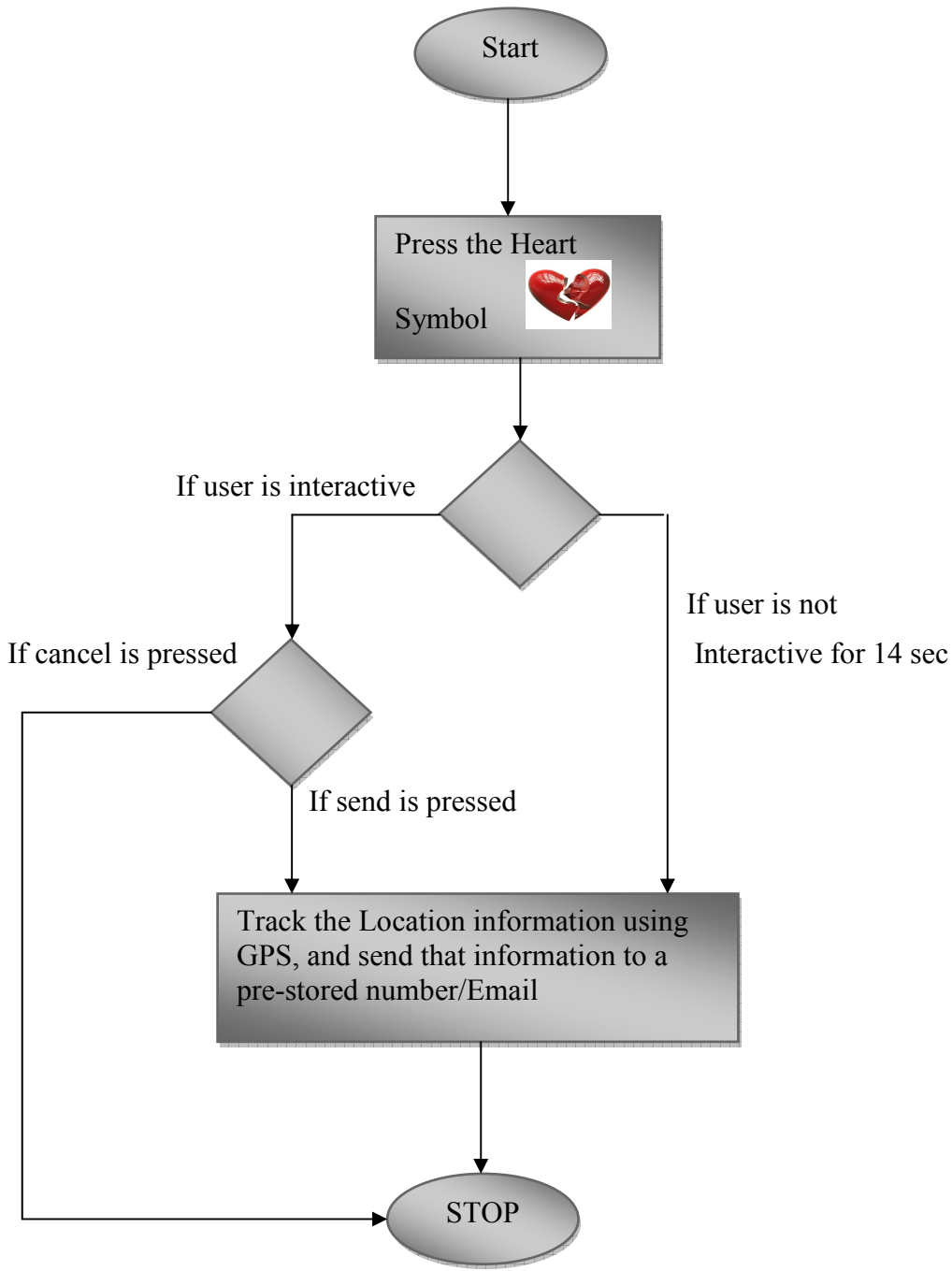

Figure 13. Implementation of Personal Safety scenario

When the Heart symbol button is pressed on the ETS home page as shown in Fig.13, the ETS application senses that user is under security risk and waits for the user to press the *send* button. In case the user is not interactive to press the button after pressing the Heart symbol, the system automatically counts down for 14 seconds and then starts to get the location information which contains the latitude and longitude coordinates using a GPS API. Subsequently, the latitude and longitude coordinates are converted into city and country location information as shown in *figure 16*, this is sent via SMS, email and along with the message "This person is under emergency take necessary action" will be posted on Facebook wall of the registered user. The entire workflow is depicted in Fig.14.

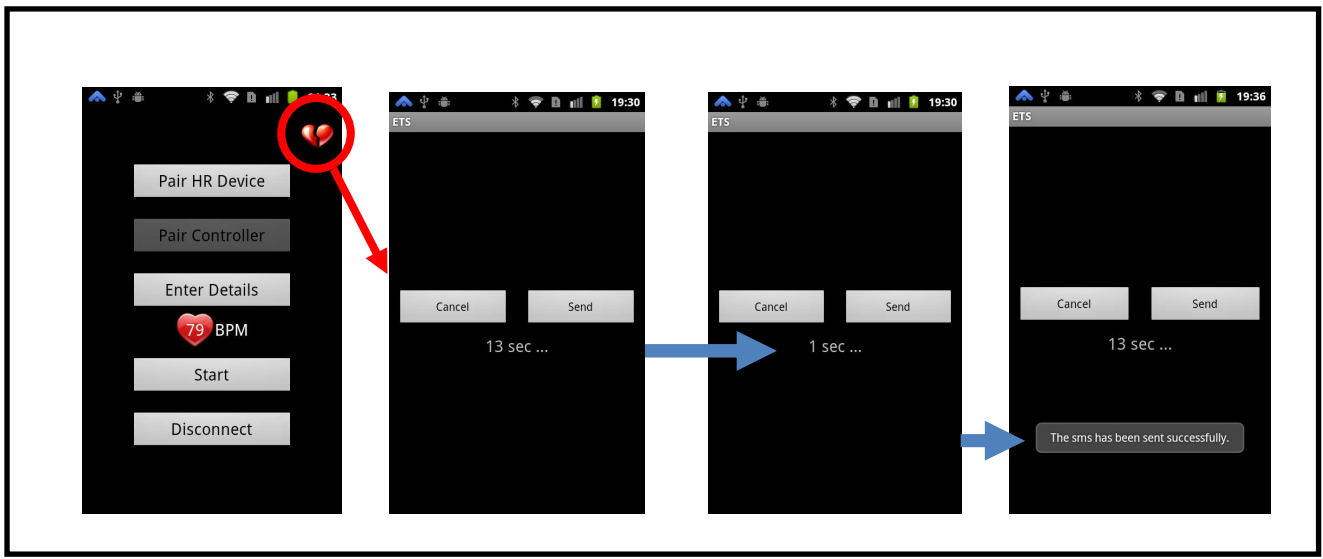

Figure 14.Workflow of Personal Safety scenario

### 4.3 Accidents

#### 4.3.1 Problem Statement

In present world, driving a vehicle has to be done with at most care, otherwise it would result in accident and the reasons behind it could be like negligent driving, drunkard driving, emergency conditions like during heart attacks and so on. The main reason for a person's death during accidents is unavailability of the first aid provision which is due to the delay in the information being reached to the hospital about the accident. The worst case to this problem is the accident occurred while a person is travelling in a vehicle. So how could we track the accident of a person as soon as it has occurred? An in built sensor of the mobile called Accelerometer would help to detect any slight movement of an object and the object being in our case is the mobile phone which is docked inside the car and not being held in hand or in pocket of the person who is driving the vehicle.

#### 4.3.2 Scenario

The design of the application is for the purpose of a person who is driving a vehicle. The scenario is, when a person is driving with the mobile phone kept in the car at a fixed position. The mobile should not be kept in pocket or held in hand of the person. Any slight movement of the mobile device would be detected by an in built sensor called Accelerometer.

#### 4.3.3 Modelling

Our application is designed using the sensor accelerometer which will detect any tilt in the mobile device. At times this tilt might be just by mistake where the person might have pressed it accidentally. In such a situation the application will wait for 14sec for the user to enter send or press cancel. If the user doesn't press any key within that time, then the application considers the person to be in danger and sends an emergency alert message to a pre stored number in the mobile. The key assumption of this application is that the mobile phone should not be kept along with the person who is driving the vehicle. This scenario can be represented using a flowchart as in Fig.15.

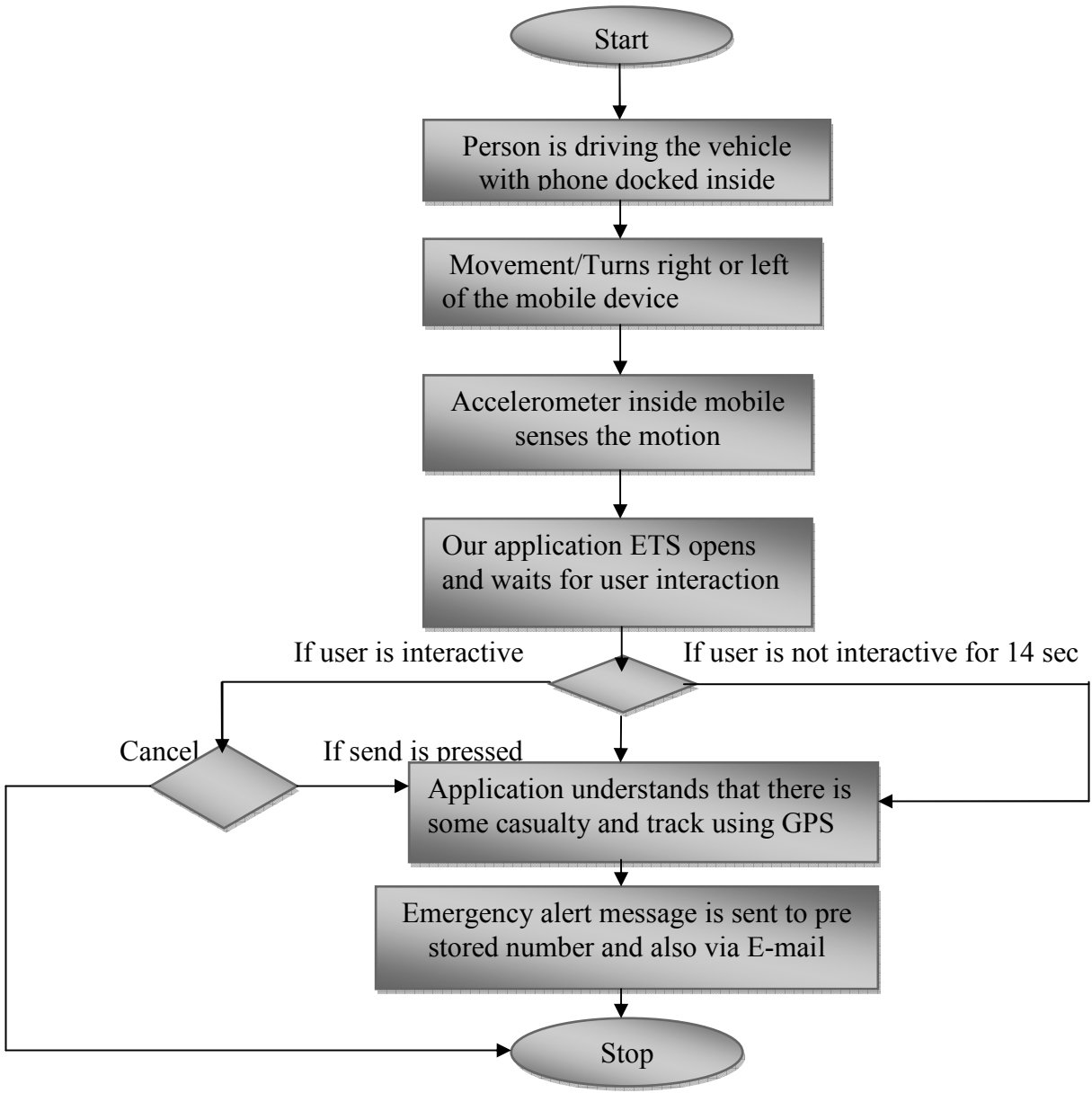

Figure 15.Modeling of Accident scenario

### 4.3.4 Implementation and Validation

Accidents, which will be sensed by the mobile using a specialized sensor called as accelerometer which is present inside android mobiles. The sensors of these type cadences the acceleration applied to the device. Reckons the device's orientation grounded as shown in Fig.16 on the rotation matrix. *SensorManager* is a public class which lets us to access the device's sensors. Below snippet is the class overview of initializing and activating the accelerometer sensor.

```
public class SensorActivity extends Activity, implements SensorEventListener
{
    private final SensorManager mSensorManager;
    private final Sensor mAccelerometer;

    public SensorActivity() {
        mSensorManager =(SensorManager)getSystemService(SENSOR_SERVICE);
        mAccelerometer =
mSensorManager.getDefaultSensor(Sensor.TYPE_ACCELEROMETER);
    }
```

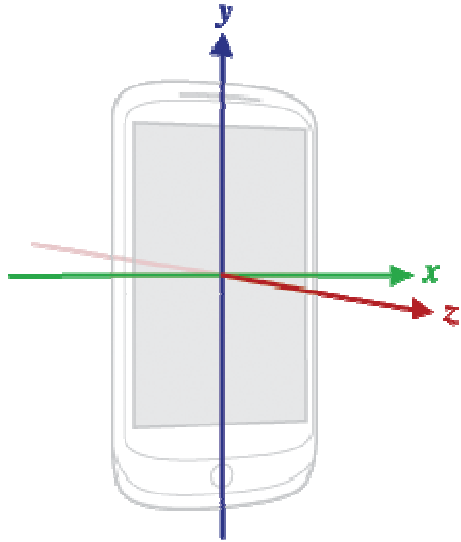

Figure 16.Movement of Mobile

The validation of the accelerometer is performed by tilting the mobile left or right or free fall motion. As soon as the there is a tilt or the mobile orientation then ETS application senses that there is a risk situation and starts to perform the operation of tracking the location information of the user before that, the ETS waits for 14 sec for user to be interactive, where user can perform 'cancel' and send as shown in Fig. 14. If there is no user interaction then automatically application sends a SMS, Email, and message is posted on Facebook wall as shown in Fig. 11.

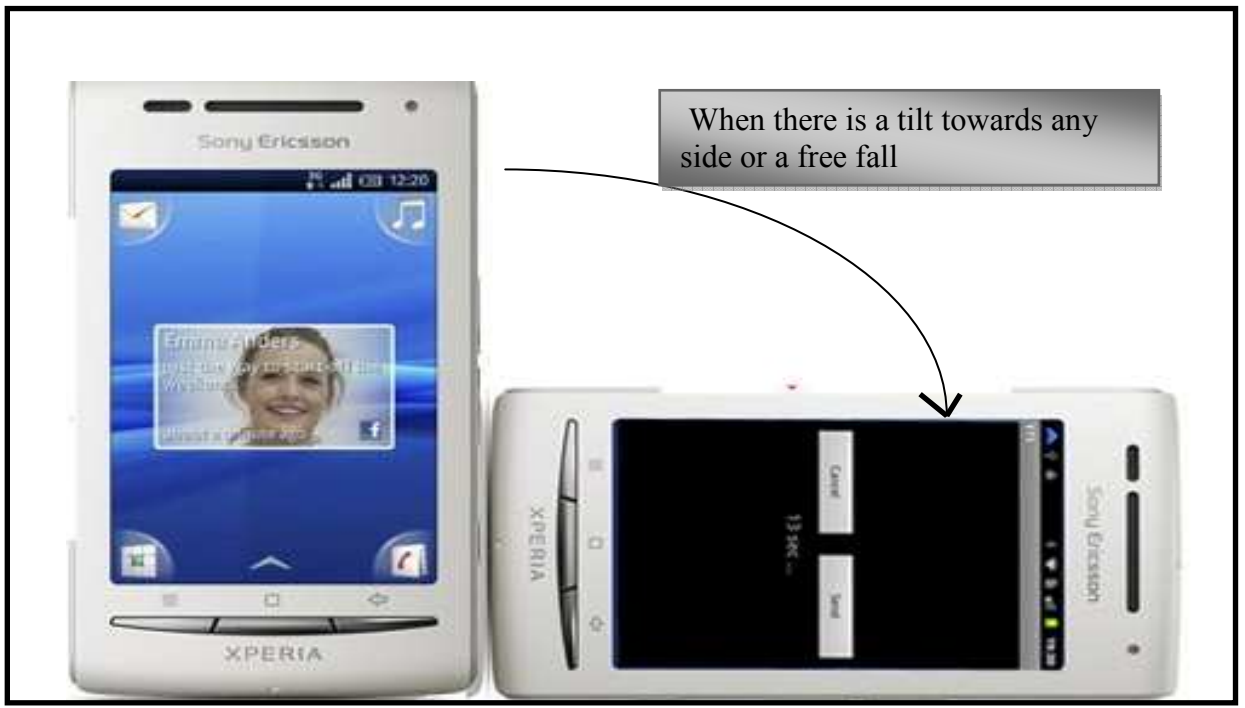

Figure 17.Sensing the movement of mobile using Accelerometer

# 5. RESULT

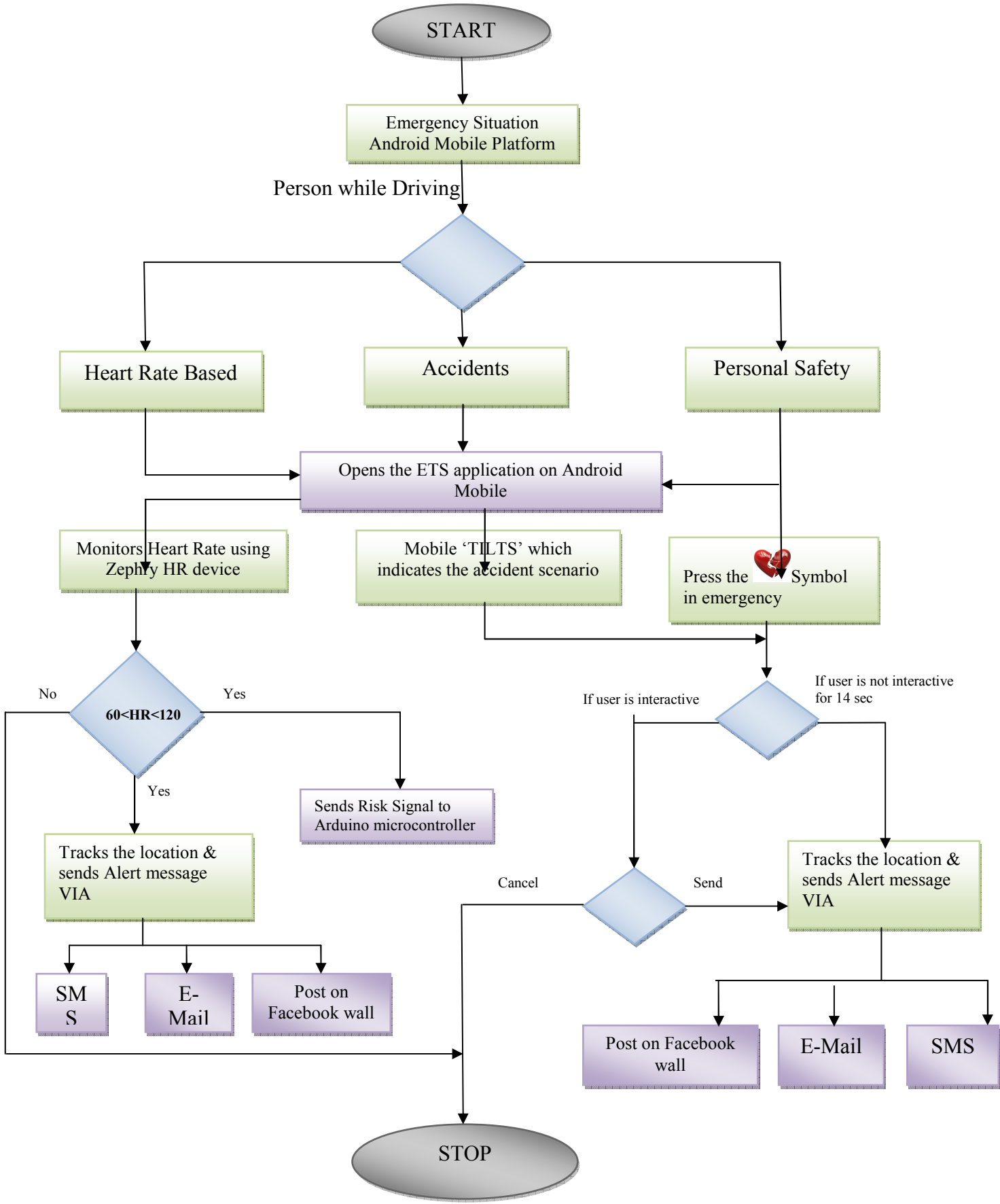

Figure 18. Working and Result of ETS Application

# CONCLUSION

Our research which was based on Emergency Tracking System using Google's Android Mobile platform and application based on that can achieve all the services and process as show in Fig.11. Emergency situation was well sensed by the Android mobile with regards to the heart rate based, accidents and personal safety. Each individual emergency scenario was researched, designed and developed. Person's heart beat was monitored using a specialized HR device which sends the heart beat rate to the mobile which in turn makes a decision with regards to the abnormal heart rate and sends an alert signal to the arduino microcontroller, simultaneously a message was sent to a pre-stored number, email address and was posted on Facebook wall successfully.

Similarly accident based emergency scenario and Personal safety can be incorporated and alert message which contains the GPS location information be sent via SMS, email and message was successfully posted on respective user's Facebook wall. Hence, Android once again proved to be a versatile operating system which allowed us to manipulate various inbuilt features of an Android mobile which made us to develop an intelligent application called as ETS.

## ACKNOWLEDGEMENTS

We would like to thank each and everyone who had helped us to go ahead with this research successfully. We would like to thank our Prof. Abbas Mohammed for being very encouraging and supportive throughout which helped us to complete the thesis very smoothly.We would like to thank our family and friends for being supportive with us all through our work.

**Authors**

Ashokkumar Ramalingam was born at Panruti, Tamilnadu in 1987. He completed his M.Sc in telecommunication engineering in Sweden and bachelor of electronics and communication engineering in India. He is young and enthusiastic person. He interests in Telecommunication, Research and Development, Wireless communication, Verification & Validation engineer and Project Management. He spends his free time with meeting friends, write blog, and play cricket, chess. He has entrepreneur and leadership skills.

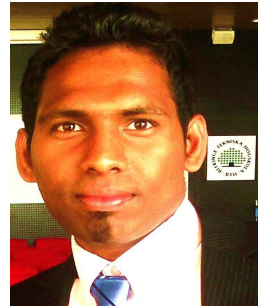

Prabhu Dorairaj was born in Chennai, India in 1985.He completed his M.S in telecommunication from Blekinge institute of technology, sweden and his Bachelors was in Electronics and communication Engineering, from Anna University,Chennai, India. He was a senior lecturer at Magna College of Engineering India from August 2007 to August 2008 and was a Software Engineer at Indus Net Technologies, India from August 2008 to February 2009.Area of interest is Mobile communication, Mobile Application testing, Software Testing and wireless Communication

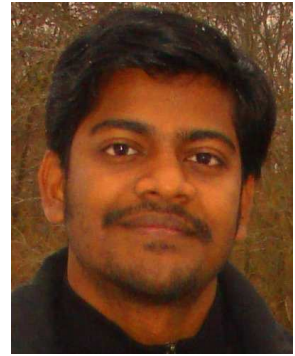

Saranya Ramamoorthy was born at Kakinada , Andhra Pradesh in 1985.She is currently working as Assistant Professor at Vignan's Institute of technology for women, Vishakapatnam, India. She pursued her post graduation in M.tech.She holds a Double Degree of both M.tech and M.S with specialization in Telecommunication Systems. She also has industrial experience of 2years 3months in TCS.Her area of interest are Mobile communication, Embedded Systems and Wireless communication.

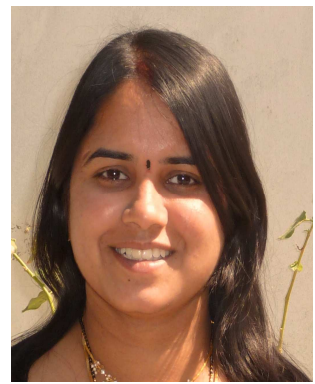